\documentstyle[12pt]{article} 
 
\begin{document}
\mbox{\hspace*{58ex}} OCU-PHYS-166 

\mbox{\hspace*{61ex}} September 1997 

\vspace*{10mm}

\begin{center}
{\Large {\bf Renormalization of number density in nonequilibrium 
quantum-field theory and absence of pinch singularities }} 
\end{center}

\hspace*{3ex}

\hspace*{3ex}

\begin{center} 
{\large {\sc A. Ni\'{e}gawa}\footnote{ 
E-mail: niegawa@sci.osaka-cu.ac.jp}

{\normalsize\em Department of Physics, Osaka City University } \\ 
{\normalsize\em Sumiyoshi-ku, Osaka 558, Japan} } \\
\end{center} 

\hspace*{2ex}

\hspace*{2ex}
\begin{center} 
{\large {\bf Abstract}} \\ 
\end{center} 
\begin{quotation}
Through introducing a notion of renormalization of particle-number 
density, a simple perturbation scheme of nonequilibrium 
quantum-field theory is proposed. In terms of the renormalized 
particle-distribution functions, which characterize the system, the 
structure of the scheme (and then also the structure of amplitudes 
and reaction rates) are the same as in the equilibrium thermal field 
theory. Then, as an obvious consequence, the amplitudes and reaction 
rates computed in this scheme are free from pinch singularities due 
to multiple products of $\delta$-functions, which inevitably present 
in traditional perturbation scheme. 
\end{quotation} 
\newpage
\setcounter{equation}{0}
Since late fifties, efforts have been made to incorporate quantum 
field theory with nonequilibrium statistical mechanics 
\cite{sch,chou,lan}. In connection with the description of 
quark-gluon plasma, which is expected to be produced in heavy-ion 
collisions and to have existed in the early universe, much work has 
recently been devoted to this issue (see, e.g., 
\cite{le-b,ume,law,alt,alt1,bai}). 

The standard framework of nonequilibrium theory is formulated by 
employing the closed-time path in a complex-time plane. A few years 
ago, Altherr and Deibert \cite{alt} have pointed out that, when 
calculating amplitudes or physical quantities in perturbative 
closed-time-path formalism, the \lq\lq pathological terms'' 
\cite{lan,le-b} necessarily appear. More precisely, such terms 
appear in self-energy-parts-inserted two-point functions in 
association with pinch singularities coming from multiple products 
of $\delta$-functions. These terms do not cancel each other unless 
the particle distributions are those for a system in thermal and 
chemical equilibrium. Since then it is shown that classes of such 
pathological terms can be resummed \cite{alt1} (see also 
\cite{chou}). An application of this result to the rate of 
hard-photon production from a nonequilibrium quark-gluon plasma 
system is made in \cite{bai}. 

In this letter, introducing a notion of renormalization of number 
density, we propose a simple perturbation scheme of quantum field 
theory for computing the rates of reactions taking place in a 
quasiuniform system near equilibrium or nonequilibrium 
quasistationary system. The scheme is the same in structure as the 
equilibrium thermal field theory (ETFT), provided that the 
equilibrium number density $n (p_0)$ in ETFT is replaced by the 
renormalized nonequilibrium number density $N (P)$ that 
characterizes the system under consideration. Then, in contrast to 
the canonical perturbation scheme, no \lq\lq pathological terms'' 
mentioned above appear in amplitudes and reaction rates evaluated 
within the scheme. 

We employ the closed-time-path formalism \cite{chou,lan,le-b}. In 
this formalism, propagators, vertices and self-energy parts enjoy 
$2 \times 2$ matrix structure. We use $\hat{B}$ to denote the 
$2 \times 2$ matrix whose $(i \, j)$ component is $B_{i j}$. A 
vertex matrix $\hat{V}$ has simple structure, 
$V_{1 2} = V_{2 1} = 0$ and $V_{1 1} = - V_{2 2}$, with $V_{1 1}$ 
the vertex factor in vacuum theory. Let $\hat{\Delta} (x_1, x_2)$ 
and $\hat{\Sigma} (x_1, x_2)$ be, in respective order, the bare 
propagator and the self-energy part in a configuration space. 
Following \cite{chou}, we assume that reactions taking place in a 
system under consideration are \lq\lq described'' by the relative or 
microscopic coordinates $x = x_1 - x_2$. Then, we make Fourier 
transforms with respect to $x^\mu$, $\hat{\Delta} (P, X)$ and 
$\hat{\Sigma} (P, X)$, where $X = (x_1 + x_2) / 2$ is the 
center-of-mass or macroscopic coordinates. It is  assumed 
\cite{chou} that the dependence of $\hat{\Delta} (P, X)$ and 
$\hat{\Sigma} (P, X)$ upon $X^\mu$ is weak. More precisely, over the 
macroscopic space-time region of the system, where a microscopic or 
elementary reaction takes place, $X^\mu$-dependence of 
$\hat{\Delta} (P, X)$ and $\hat{\Sigma} (P, X)$ may be ignored. 
Then, in calculating the reaction rate, we can use 
$\hat{\Delta} (P, X)$ and $\hat{\Sigma} (P, X)$ with fixed $X^\mu$. 
This is the first stage of the theoretical analysis. Microscopic 
reactions cause changes in the number densities of (quasi)particles, 
through which the density matrix changes with macroscopic space-time 
$X^\mu$. Dealing with this is the subject of the second stage, where 
(weak) $X^\mu$-dependence of $\hat{\Delta} (P, X)$ and 
$\hat{\Sigma} (P, X)$ are explicitly taken into account. In this 
letter, as in \cite{alt,alt1,bai}, we concentrate our concern on the 
first stage and drop the argument $X$ throughout. 

The matrix elements of $\hat{\Delta} (P)$ and $\hat{\Sigma} (P)$ 
enjoy various properties \cite{chou}. Among those we shall use 
$Re \Sigma_{1 2} = Re \Sigma_{2 1} = 0$, 
$\Sigma_{2 2} = - \Sigma^{\, *}_{1 1}$, 
$\sum_{i, \, j = 1}^2 (-)^{i + j} \Delta_{i \, j} (P) = 0$ and 
$\sum_{i, \, j = 1}^2 \Sigma_{i j} (P) = 0$. 

For simplicity of presentation, we take a massless, 
self-interacting, complex-scalar field theory with a conserved 
charge. $\hat{\Delta} (P)$ may be written as 
\begin{equation} 
\hat{\Delta} (P) = \hat{M} (P) \, \hat{\Delta}_F (P) \, \hat{M} (P) 
\, , 
\label{prop} 
\end{equation} 
where\footnote{
In calculating an amplitude in ETFT, a common practice \cite{lan} is 
to keep $\epsilon$ finite throughout and at the end of calculation 
the limit $\epsilon \to 0^+$ is taken. We follow this procedure.} 
\begin{eqnarray} 
\hat{\Delta}_F (P) & = & \mbox{diag} \left( \Delta_F (P), \, - 
\Delta_F^* (P) \right) \;\;\;\;\;\;\;\;\;\; \left( \Delta_F (P) = 
1 / (P^2 + i \epsilon) \right) \, , 
\label{diag} \\ 
\hat{M} (P) & = & \left( 
\begin{array}{cc}  
\sqrt{1 + n (P)} & \frac{\theta (- p_0) + n (P)}{\sqrt{1 + n (P)}} 
\\ 
\frac{\theta (p_0) + n (P)}{\sqrt{1 + n (P)}} & \sqrt{1 + n (P)} 
\end{array} 
\right) \, . 
\label{bog} 
\end{eqnarray} 
Here $n (P) = n (p_0, \, p)$ with $p = |{\bf p}|$ is the 
number-density function that characterizes the ensemble of the 
systems. In the case where the \lq\lq local'' temperature 
$T = \beta^{- 1}$ and the chemical potential $\mu$, being conjugate 
to the charge, are defined, 
$n (P) = 1 / [ e^{\beta (|p_0| - \epsilon (p_0) \mu)} - 1]$ with 
$\beta$ and $\mu$ the functions of macroscopic space-time 
coordinates. 

Here we recall that, in ETFT, the self-energy-part matrix 
$\hat{\Sigma} (P)$ takes \cite{lan} the form 
\begin{equation} 
\hat{\Sigma} (P) = \hat{M}^{- 1} (P) \, \hat{\Sigma}_F (P) \, 
\hat{M}^{- 1} (P) \;\;\;\;\;\;\; \hat{\Sigma}_F (P)  = \mbox{diag} 
\left( \Sigma_F (P) \, - \Sigma_F^* (P) \right) \, . 
\label{equi} 
\end{equation} 
One can easily see from Eqs. (\ref{prop}), (\ref{diag}) and 
(\ref{equi}) that an $l$ $(\geq 1)$ self-energy-parts-inserted 
two-point function $\hat{G}_l (P) \equiv \hat{\Delta} (P) 
[ \hat{\Sigma} (P) \, \hat{\Delta} (P) ]^l$ does not include 
functions 
$(P^2 + i \epsilon)^{- k} (P^2 - i \epsilon)^{- (l + 1 - k)}$ 
$(1 \leq k \leq l)$ but includes functions 
$(P^2 \pm i \epsilon)^{- (l + 1)}$. In the limit $\epsilon \to 0^+$, 
the former functions develop pinch singularity while the latter 
functions turn out to the well-defined distributions, and thus 
$\hat{G}_l (P)$ is free from pinch singularity. In the present 
nonequilibrium case, Eq. (\ref{equi}) does not hold and 
$\hat{G}_l (P)$ includes \lq\lq pinch-singular'' functions 
$(P^2 + i \epsilon)^{- k} (P^2 - i \epsilon)^{- (l + 1 - k)}$. This 
is essentially what has been observed in \cite{alt}. 

Let us construct the renormalization theory by introducing a 
renormalized or \lq\lq physical'' number density $N (P)$, 
\begin{equation} 
N (P) = n (P) + \delta n (P)\, , 
\label{renorm} 
\end{equation} 
with which the bare (renormalized) propagator reads (cf. Eqs. 
(\ref{prop}) - (\ref{bog})) 
\begin{eqnarray} 
\hat{\Delta}^{(r)} (P) & = & \hat{\Delta} (P) \, 
\rule[-3mm]{.14mm}{8mm} \raisebox{-2.85mm}{\scriptsize{$\; n (P) 
\to N (P)$}} \, , \nonumber \\  
\hat{M}^{(r)} (P) & = & \hat{M} (P) \, 
\rule[-3mm]{.14mm}{8mm} \raisebox{-2.85mm}{\scriptsize{$\; n (P) 
\to N (P)$}} \, .  
\label{ren} 
\end{eqnarray} 
In order to compensate the difference between the renormalized 
$\hat{\Delta}^{(r)}$ and the original \lq\lq bare'' $\hat{\Delta}$, 
$\delta \hat{\Delta} (P)$ $\equiv$ $\hat{\Delta}^{(r)} (P)$ $-$ 
$\hat{\Delta} (P)$, we should introduce the counter term in the 
interaction Lagrangian, 
\begin{equation} 
L_c = - \frac{1}{2} \int \prod_{\ell = 1}^4 \left( d^{\, 4} x_\ell 
\right) \sum_{i, \, j = 1}^2 \phi_i^* (x_1) 
\left[ \hat{\Delta}^{- 1} (x_1, \, x_2) \, \delta 
\hat{\Delta} (x_2, x_3) \, \hat{\Delta}^{- 1} (x_3, \, x_4) 
\right]_{i j} \phi_j (x_4) \, , 
\label{lc} 
\end{equation} 
where $\phi_j$ and $\phi_j^*$ ($j = 1, 2$) stand for the type-$j$ 
fields and 
\begin{eqnarray} 
\hat{\Delta}^{- 1} (x, y) & \equiv & \int 
\frac{d^{\, 4} P}{(2 \pi)^4} \, e^{- i P \cdot (x - y)} \, 
\hat{\Delta}^{- 1} (P) \nonumber \\ 
\delta \hat{\Delta} (x, y) & \equiv & \int 
\frac{d^{\, 4} P}{(2 \pi)^4} \, e^{- i P \cdot (x - y)} \, \delta 
\hat{\Delta} (P) \nonumber \\ 
& = & \int \frac{d^{\, 4} P}{(2 \pi)^4} \, e^{- i P \cdot (x - y)} 
\left[ - 2 \pi i \, \delta n (P) \, \delta_\epsilon (P^2) \, 
\hat{A} \right] \, . 
\label{del-D} 
\end{eqnarray} 
Here $A_{i j}$ $= 1$ $(i, \, j = 1, \, 2)$ for all $i$ and $j$ 
and$^{\dagger}$ $\delta_\epsilon (P^2) \equiv \epsilon / 
[ \pi \{ (P^2)^2 + \epsilon^2 \}]$. 

The perturbative calculation goes with propagator 
$\hat{\Delta}^{(r)} (P)$ in Eq. (\ref{ren}) and \lq\lq vertices'' 
coming from $L_{int} + L_c$, where $L_{int}$ is the original 
interaction Lagrangian. Conforming to the ultra-violet 
renormalization theory, one can say that $n (P)$ is the bare 
number-density function and $N (P)$ is the renormalized or physical 
number-density function. The latter describes the physical system 
under consideration. The self-energy part $\hat{\Sigma}^{(r)} (P)$ 
obtained from the above perturbation scheme takes the 
form\footnote{ 
Suppose that we are analyzing $\hat{\Sigma}^{(r)}$ in the $n$th 
order of perturbation series. The second term on the right-hand side 
of Eq. (\ref{ren-con1}) comes from $L_c$ with the $n$th order 
counter term, being proportional to $\delta n^{(n)} (P)$. The 
$n$th order $\hat{\Sigma} (P)$ in Eq. (\ref{ren-con1}) involves 
$\delta n^{(j)} (P)$ with $1 \leq j \leq n - 1$.} 
\begin{equation} 
\hat{\Sigma}^{(r)} (P) = \hat{\Sigma} (P) - \hat{\Delta}^{- 1} (P) 
\, \delta \hat{\Delta} (P) \, \hat{\Delta}^{- 1} (P) \, . 
\label{ren-con1} 
\end{equation} 

We are now in a position to set up the renormalization condition: 
{\em The self-energy part} $\hat{\Sigma}^{(r)} (P)$ {\em is the same 
in structure as Eq. (\ref{equi}) in ETFT}, i.e., 
\begin{equation} 
\hat{\Sigma}^{(r)} (P) = [\hat{M}^{(r)} (P)]^{- 1} \, 
\hat{\Sigma}_{F}^{(r)} (P) \, [\hat{M}^{(r)} (P)]^{- 1} 
\label{self} 
\end{equation} 
where 
\begin{equation} 
\hat{\Sigma}_F^{(r)} (P) = \mbox{diag} \left( \Sigma^{(r)}_F (P), \, 
- (\Sigma^{(r)}_F (P))^* \right) \, . 
\label{diag1} 
\end{equation} 

The second term on the right-hand side of Eq. (\ref{ren-con1}) does 
not contribute to $(\Sigma_F^{(r)})_{1 1}$ and 
$(\Sigma_F^{(r)})_{2 2}$ and the condition $[\{ \Sigma_F^{(r)} (P) 
\}_{2 2}]^* = - \{ \Sigma_F^{(r)} (P) \}_{1 1}$ is automatically 
met; 
\begin{equation} 
\Sigma_F^{(r)} (P) = \Sigma_{1 1} (P) + \theta (p_0) \, \Sigma_{1 2} 
(P) + \theta (- p_0) \, \Sigma_{2 1} (P) \, . 
\label{sigma} 
\end{equation} 
From Eq. (\ref{ren}) with Eq. (\ref{bog}) and Eqs. (\ref{del-D}) and 
(\ref{ren-con1}), we see that the condition 
$\{ \Sigma_F^{(r)} (P) \}_{1 2} = \{ \Sigma_F^{(r)} (P) \}_{2 1} 
= 0$ yields 
\begin{eqnarray} 
\delta n (P) \, \delta_\epsilon (P^2) & = & - \frac{i}{2 \pi} \, 
\frac{1}{P^2 - i \epsilon} \, \frac{1}{P^2 + i \epsilon} \left[ 
\theta (p_0) \left\{ (1 + N (P)) \, \Sigma_{1 2} (P) - N (P) \, 
\Sigma_{2 1} (P) \right\} \right. \nonumber \\ 
& & \left. + \theta (- p_0) \left\{ (1 + N (P)) \Sigma_{2 1} (P) - N 
(P)) \, \Sigma_{1 2} (P) \right\} \right] \, . 
\label{shuppatsu} 
\end{eqnarray} 
Now let us introduce 
\begin{eqnarray} 
\Gamma_{p}^{(net)} (P) & \equiv & \theta (p_0) [ \{ 1 + N (P) \} 
\Gamma_{p} (P) - N (P) \Gamma_{d} (P) ] \nonumber \\ 
& & + \theta (- p_0) [ \{1 + N (P) \} \overline{\Gamma}_{p} (P) - N 
(P) \overline{\Gamma}_{d} (P) ] \, , 
\label{net} 
\end{eqnarray} 
where 
\begin{eqnarray} 
\Gamma_{p} (P) & = & \overline{\Gamma}_{d} (P) = - \frac{i}{2 p} \, 
\Sigma_{1 2} (P) \, , \nonumber \\ 
\Gamma_{d} (P) & = & \overline{\Gamma}_{p} (P) = - \frac{i}{2 p} \, 
\Sigma_{2 1} (P) \, . 
\label{iwa} 
\end{eqnarray} 
On the mass shell $p_0 = p$ ($p_0 = - p$), $\Gamma_p^{(net)} (P)$ is 
the net production rate of a particle (antiparticle) with momentum 
${\bf p}$ ($- {\bf p}$) \cite{chou}. Then $\Gamma_p^{(net)} (P)$ in 
Eq. (\ref{net}) is an off-shell extension of the net production 
rate. Substitution of Eq. (\ref{net}) with Eq. (\ref{iwa}) into Eq. 
(\ref{shuppatsu}) yields 
\begin{equation} 
\delta n (P) \, \delta_\epsilon (P^2) = \frac{p}{\pi} \, 
\frac{1}{(P^2)^2 + \epsilon^2} \, \Gamma^{(net)}_p (P) \, . 
\label{kettei} 
\end{equation} 

It is to be noted in passing that, in ETFT, $\Gamma^{(net)}_p (P)$ 
vanishes (see, e.g., \cite{chou}). 

In the limit $\epsilon \to 0^+$, $\delta n (P)$ in Eq. 
(\ref{kettei}) is singular due to the pinch singularity at 
$|p_0| = p$. Equation (\ref{kettei}) may be \lq\lq solved'' for 
$\delta n (P)$ in the form, 
\label{hito} 
\begin{eqnarray} 
\delta n (P) & = & 2 (p^2 + \tilde{\epsilon}^2) \, 
\Gamma_{p}^{(net)} (P) \int_{- \infty}^{+ \infty} 
\frac{d k_0}{2 \pi} \, \Delta_F^* (k_0, \, p) \, 
\Delta_F (k_0, \, p) \label{sing} \\ 
& = & \frac{p}{\epsilon} \, \Gamma_{p}^{(net)} (P) \, , 
\label{formal} 
\end{eqnarray} 
where $\tilde{\epsilon} \simeq \epsilon / 2 p$. 

It is obvious from the above construction that the renormalization 
theory is obtained from the ETFT simply by substituting $N (P)$ for 
the equilibrium distribution function $n_B (p_0)$ 
$(= 1 / [e^{\beta (|p_0| - \epsilon (p_0) \mu)} - 1])$. Then, the 
fact in ETFT that a self-energy-parts-inserted propagator is free 
from pinch singularities (cf. above after Eq. (\ref{equi})) is 
transmitted as it is to the case of renormalization theory. In fact, 
from Eq. (\ref{ren}) with Eqs. (\ref{prop}) and (\ref{diag}), and 
Eqs. (\ref{self}) and (\ref{diag1}), we see that $\hat{\Delta}^{(r)} 
(P) [ \hat{\Sigma}^{(r)} (P) \, \hat{\Delta}^{(r)} (P) ]^l$ 
$(l \geq 1)$ includes only well-defined function 
$(P^2 \pm i \epsilon)^{- (l + 1)}$. 

We now demonstrate that, as far as the {\em singular pieces} are 
concerned, $n (P)$ on the mass shell $p_0 = p$ [$p_0$ $=$ $- p$] 
coincides with the Heisenberg particle [antiparticle] number density 
$n_a^H (p)$ [$n_b^H (p)$]. $n_a^H (p)$ reads, with obvious notation, 
\begin{eqnarray} 
\langle a^\dagger ({\bf p}, \, t) \, a ({\bf p}', \, t) \rangle 
& \equiv & \delta ({\bf p} - {\bf p}') \, n_a^H (p) \nonumber \\ 
& = & \frac{1}{2 p} \int \frac{d^{\, 3} x \, d^{\, 3} y}{(2 \pi)^3} 
\, e^{-i (P \cdot x - P' \cdot y)} 
\stackrel{\leftrightarrow}{\frac{\partial}{\partial x_0}} 
\, \stackrel{\leftrightarrow}{\frac{\partial}{\partial y_0}} \, 
\langle \phi^* (x) \, \phi (y) \rangle \, 
\rule[-3mm]{.14mm}{8.5mm} \raisebox{-2.85mm}{\scriptsize{$\; x_0 = 
y_0 = t$}} \, , \nonumber \\ 
\label{part} 
\end{eqnarray} 
where $P = (p, {\bf p})$ and $P' = (p', {\bf p}')$. For the case of 
antiparticle number density, we have Eq. (\ref{part}) with 
$\phi^* \leftrightarrow \phi$. Note that 
$\langle \phi^* (x) \, \phi (y) \rangle$ 
[$\langle \phi (x) \, \phi^* (y) \rangle$] is the $(1 2)$ [$(2 1)$] 
component \cite{chou,lan,le-b} of the full propagator 
$i G_{1 2} (y - x)$ [$i G_{2 1} (x - y)$], we obtain 
\begin{equation} 
n_{a / b}^H (p) = \frac{1}{2 p} \int_{- \infty}^{+ \infty} 
\frac{d k_0}{2 \pi} \, (k_0 \pm p)^2 \, i G_{1 2 / 2 1} 
(k_0, \, p) \, . 
\end{equation} 

We compute $n_{a / b}^H (p)$ up to the contribution from one 
self-energy-part-inserted $G_{1 2 / 2 1} (k_0, \, p)$. We first 
employ the perturbation scheme in the renormalization theory defined 
above. As mentioned above, computation goes just as in the case of 
ETFT. The resultant $n_{a / b}^H (p)$ reads 
\begin{equation} 
n_{a / b}^H (p) = N (\pm p, \, p) + \delta N (\pm p, \, p) \, . 
\label{new} 
\end{equation} 
Here $N$ comes from the lowest-order contribution to $G_{1 2 / 2 1}$ 
and $\delta N$ comes from the one self-energy-part-inserted 
$G_{1 2 / 2 1}$. As in the case of ETFT, $\delta N$ is a 
well-defined at most finite functional of $N$. 

Now we turn to compute $n_{a / b}^H (p)$ in {\em traditional 
scheme}, without introducing the renormalization counter term. The 
lowest-order contribution to $G_{1 2 / 2 1}$ yields 
$n_{a / b}^H (p)$ $=$ $n (\pm p, \, p)$, where $n$ is the 
\lq\lq starting'' number density function in Eq. (\ref{bog}). Then, 
we write 
\begin{equation} 
n_{a / b}^H (p) = n (\pm p, \, p) + \delta n_{a / b}^H (p) \, . 
\label{old} 
\end{equation} 
The contribution from one self-energy-part-inserted $G_{1 2 / 2 1}$ 
to $\delta n_{ a / b}^H (p)$ reads, 
\begin{eqnarray} 
\delta n_{a / b}^H (p) & = & \int_{- \infty}^{+ \infty} 
\frac{d k_0}{2 \pi} \, (k_0 \pm p)^2 \left[ \Delta_F^* \, \Delta_F 
\, \tilde{\Gamma}_p^{(net)} (k_0, \, p) \right. \nonumber \\ 
& & \left. - (\theta_\mp + n) \left\{ (\Delta_F)^2 \tilde{\Sigma}_F 
- (\Delta_F^*)^2 \tilde{\Sigma}_F^* \right\} \right] \, , 
\label{yaw-1} 
\end{eqnarray} 
where $\tilde{\Sigma}_F = \tilde{\Sigma}_{1 1} + \theta_+ 
\tilde{\Sigma}_{1 2} + \theta_- \tilde{\Sigma}_{2 1}$, $\theta_\pm 
\equiv \theta (\pm k_0)$, $n = n (k_0, \, p)$, 
$\Delta_F = \Delta_F (k_0, \, p)$ and 
$\tilde{\Sigma}_F = \tilde{\Sigma}_F (k_0, \, p)$. 
$\tilde{\Gamma}_p^{(net)} (k_0, p)$ in Eq. (\ref{yaw-1}) is the 
traditional-scheme counterpart of $\Gamma_p^{(net)} (P)$ in Eq. 
(\ref{net}) with Eq. (\ref{iwa}). In the limit $\epsilon \to 0^+$, 
the term with $\Delta_F^* \, \Delta_F$ in Eq. (\ref{yaw-1}) diverges 
due to pinch singularity at $|k_0| = p$. On the other hand, the 
contributions from $(\Delta_F)^2$ and $(\Delta^*_F)^2$, being 
well-defined functions, are at most finite. We deal only with the 
singular contribution coming from $|k_0| = p$, 
\begin{equation} 
\delta n_{a / b}^H (p) \simeq 2 p^2 \, \tilde{\Gamma}_p^{(net)} 
(\pm p, \, p) \int_{- \infty}^{+ \infty} \frac{d k_0}{2 \pi} 
\Delta_F^* (k_0, \, p) \, \Delta_F (k_0, \, p) \, , 
\label{yaw-2} 
\end{equation} 
where \lq $\simeq$' is used to denote an approximation that is valid 
for keeping the singular contribution. In obtaining Eq. 
(\ref{yaw-2}), use has been made of the fact that $\Delta_F^* \, 
\Delta_F$ is even function of $k_0$. $\delta n (P)$ on the mass 
shell $p_0 = \pm p$, Eq. (\ref{sing}), and the singular piece of 
$\delta n_{a / b}^H (p)$ in Eq. (\ref{yaw-2}) is the same in form. 
It should be noted, however, that $\tilde{\Gamma}_p^{(net)}$ $\neq$ 
$\Gamma_p^{(net)}$. We rewrite $\tilde{\Gamma}_p^{(net)}$ in terms 
of $N$ by using 
$n (\pm p, \, p) = N (\pm p, \, p) + \{ n (\pm p, \, p) - N 
(\pm p, \, p) \}$. The philosophy of renormalization tells us that 
the difference $n - N$ only affects the higher order contribution to 
$n_{a / b}^H (p)$. Thus, to the accuracy of one self-energy-part 
insertion, we have 
$\delta n_{a / b}^H (p) \simeq \delta n (\pm p, \, p)$. Then, from 
Eqs. (\ref{new}) and (\ref{old}), we find 
\begin{eqnarray} 
n^H_{a / b} (p) \simeq N (\pm p, \, p) & \simeq & n (\pm p, p) + 
\delta n_{a / b}^H (p) \nonumber \\ 
& \simeq & n (\pm p, \, p) + \delta n (\pm p, \, p) \, , 
\label{hosi} 
\end{eqnarray} 
where use has been made of the fact that $\delta N$, Eq. 
(\ref{new}), is not singular. The relation (\ref{hosi}) is in accord 
with Eq. (\ref{renorm}) on the mass shell $p_0 = \pm p$. 

Thus, we have learned that, as far as the singular pieces are 
concerned, the renormalized number density $N (P)$ (Eq. 
(\ref{renorm})) on the mass shell $p_0 = \pm p$ coincides with the 
Heisenberg number density $n_{a / b}^H (p)$. Then we may regard 
$N (P)$ as an off-shell extension of the Heisenberg number density. 

In passing it is worth making a comment. The first computation 
leading to Eq. (\ref{new}) is based on the theory, in which the free 
Lagrangian is written in terms of physical or renormalized number 
density $N (P)$. On the other hand, the second computation is based 
on the theory, in which the free Lagrangian is written in terms of 
original bare number density $n (P)$. As in the ultra-violet 
renormalization theory in quantum field theories, both approaches 
are equivalent. In the first approach, the renormalization is 
\lq\lq done'' at the beginning by introducing the counter Lagrangian 
$L_c$. While the second approach starts with the bare Lagrangian and 
the renormalization is \lq\lq done'' at the end. 

Now let us turn to analyze the physical meaning of the {\em singular 
part} of $\delta n_{a / b}^H (p)$, Eq. (\ref{yaw-2}). Introducing 
the Fourier transform of $\Delta_F (k_0, \, p)$, Eq. (\ref{diag}), 
\begin{eqnarray} 
\Delta_F (t, \, p) & \equiv & \int_{- \infty}^{+ \infty} 
\frac{d k_0}{2 \pi} \, e^{- i k_0 t} \Delta_F (k_0, \, p) 
\nonumber \\ 
& = & - \frac{i}{2 p} \left[ \theta (t) e^{- i 
(k - i \tilde{\epsilon}) t } + \theta (- t) 
e^{i (k - i \tilde{\epsilon}) t } \right] \, , 
\end{eqnarray} 
we obtain 
\begin{equation} 
\int_{- \infty}^{+ \infty} \frac{d k_0}{2 \pi} \, 
\Delta_F^* (k_0, \, p) \, \Delta_F (k_0, \, p) = 
\int_{- \infty}^{+ \infty} d t \, \Delta_F^* (- t, \, p) \, 
\Delta_F (t, \, p) \, . 
\label{kawa} 
\end{equation} 
Here let us recall that, in standard Lippmann-Schwinger formalism, a 
singular function $2 \pi [\delta (E_f - E_i)]^2$ appears in a 
transition probability of some reaction, where $E_i$ ($E_f$) is the 
energy of the initial (final) state of the reaction. We follow the 
standard argument for interpreting this function as 
$(t_f - t_i) \delta (E_f - E_i)$, where $t_f - t_i$ is the time 
interval during which the interaction acts. Applying this to the 
present case, we have 
\begin{eqnarray} 
\int_{t_i}^{t_f} d t \, \Delta_F^* (- t, \, p) \, \Delta_F (t, \, p) 
& = & \frac{1}{4 p^2} \left[ 
\frac{1 - e^{2 \tilde{\epsilon} t_i}}{2 \tilde{\epsilon}} - \frac{ 
e^{- 2 \tilde{\epsilon} t_f} - 1}{2 \tilde{\epsilon}} \right] 
\label{shou} \\ 
& \simeq & \frac{1}{4 p^2} \, (t_f - t_i) \, . 
\nonumber 
\end{eqnarray} 
Substituting this into Eq. (\ref{kawa}) and then into Eq. 
(\ref{yaw-2}), we obtain 
\begin{equation} 
\delta n_{a / b}^H (p) \simeq \frac{t_f - t_i}{2} \, 
\Gamma_p^{(net)} (\pm p, \, p) \, . 
\label{ketsu}
\end{equation} 
It is to be noted in passing that if we take the limit 
$t_i \to - \infty$ and $t_f \to + \infty$ in Eq. (\ref{shou}), we 
have $1/4 p^2 \tilde{\epsilon}$ $\simeq$ $1 / 2 p \epsilon$. 
Substitution of this into Eq. (\ref{yaw-2}) reproduces Eq. 
(\ref{formal}) on the mass shell $p_0 = \pm p$. 

It is interesting to note that Eq. (\ref{ketsu}) is {\em half} of 
the net production probability during the time interval $t_f - t_i$ 
of the reaction. What the result (\ref{ketsu}) tells us is the 
following. Since $\Gamma_p^{(net)}$ $\neq 0$, the number density 
$\tilde{n}_{a / b} (p, \, t)$ changes with microscopic time $t$. At 
the initial time $t_i$ $(\sim - \infty)$ of some reaction 
$\tilde{n}_{a / b} (p, \, t_i)$ $=$ $n (\pm p, \, p)$, Eq. 
(\ref{bog}), and at the final time $t_f$ $(\sim + \infty)$ 
$\tilde{n}_{a / b} (p, t_f) = \tilde{n}_{a / b} (p, \, t_i) + 
(t_f - t_i) \Gamma_p^{(net)} (\pm p, \, p) \simeq n (\pm p, \, p) + 
2 \delta n_{a / b} (p)$. Then, $\delta n^H_{a / b} (p)$ in Eq. 
(\ref{ketsu}) is $\delta n^H_{a / b} (p) = \tilde{n}_{a / b} 
(p, (t_f + t_i) / 2) - \tilde{n}_{a / b} (p, t_i)$, i.e., 
$N (\pm p, p) \simeq n (\pm p, p) + \delta n^H_{a / b} (p)$, Eq. 
(\ref{hosi}), is the number density $\tilde{n}_{a / b} (p, t)$ at 
the middle or average time of the initial and final times of the 
reaction, $t = (t_f + t_i) / 2$. 

Here we briefly mention how to treat the gauge field and massless 
fermion field. In the case of gauge field like gluon in QCD, we 
employ the Coulomb gauge or the Landshoff-Rebhan's variant 
\cite{l-r} of the covariant gauge. The $n (P)$-dependent part of 
$\hat{\Delta} (P)$ and $\hat{\Sigma} (P)$ are decomposed into two 
sectors, the electric sector $(e)$ and the magnetic sector $(m)$. 
For each sector $\tau$ $(= e, m)$, we introduce the Bogoliubov 
matrix, which is given by Eq. (\ref{bog}) with 
$n (P) \to n^{(\tau)} (P)$. For massless fermion field, we decompose 
$\hat{\Delta} (P)$ and $\hat{\Sigma} (P)$ into two sectors, 
$\hat{\Delta} (P) = \sum_{\tau = \pm} {\cal P}_\tau 
\hat{\Delta}^{(\tau)} (P)$ and $\hat{\Sigma} (P) = 
\sum_{\tau = \pm} {\cal P}_\tau \hat{\Sigma}^{(\tau)} (P)$, where 
${\cal P}_\tau \equiv (\gamma_0 - \tau \hat{{\bf p}} \cdot 
\vec{\gamma}) / 2$. In particular, $\hat{\Delta}_F (P) = 
\sum_{\tau = \pm} {\cal P}_\tau \hat{\Delta}_F^{(\tau)} (P)$ with 
$\hat{\Delta}_F^{(\tau)} (P) = \mbox{diag} [1 / \{ p_0 
(1 + i \epsilon) - \tau p \}, - 1 / 
\{ p_0 (1 - i \epsilon) - \tau p \}]$ (cf. Eq. (\ref{diag})). For 
each sector $\tau$ $(= \pm)$, we introduce the Bogoliubov matrix, 
Eq. (\ref{bog}), with $n (P) \to - n^{(\tau)} (P)$. With this 
preliminaries, the analysis goes as in the case of complex-scalar 
field. 

Finally we mention two related work. The first one is the so-called 
nonequilibrium thermo field dynamics \cite{ume}. This is a \lq\lq 
single-time'' formalism, without distinction between the 
microscopic- and macroscopic-times. The time representation, rather 
than the $p_0$ representation, is employed and through carrying out 
renormalization of the number density a nonrelativistic 
Schr\"odinger field theory is analyzed. Since the time 
representation is used, the structure of the (Schr\"odinger) 
propagators in a $p_0$ space may not be seen directly. The second 
one \cite{law} starts with performing the approximate resummation of 
the absorptive parts of the self-energy parts in relativistic scalar 
field theories. Then, the perturbation scheme is so constructed that 
the renormalized free Lagrangian $L_0^{(r)}$ yields the 
approximately resummed propagator. The counter Lagrangian (the 
difference between the original free Lagrangian and $L_0^{(r)}$) is 
local. By contrast, the counter Lagrangian (\ref{lc}) in our 
renormalization scheme is nonlocal as in the hard-thermal-loop 
resummation scheme \cite{le-b, pis} in ETFT. 
\section*{Acknowledgments} 
This work was supported in part by the Grant-in-Aide for Scientific 
Research ((A)(1) (No. 08304024)) of the Ministry of Education, 
Science and Culture of Japan. 
 
\end{document}